\documentclass[journal,11pt,draftclsnofoot,onecolumn]{IEEEtran}
\ifCLASSINFOpdf
\usepackage[pdftex]{graphicx}
  \graphicspath{{../pdf/}{../jpeg/}}
  \DeclareGraphicsExtensions{.pdf,.jpeg,.png}
\else
  \usepackage[dvips]{graphicx}
  \graphicspath{{../eps/}}
  \DeclareGraphicsExtensions{-eps-converted-to.pdf}
\fi
\usepackage{epstopdf}
\usepackage{amsmath,amsthm, amssymb}
\usepackage{threeparttable}
\usepackage{bm}
\usepackage{multirow}
\usepackage{booktabs}
\usepackage{color}
\usepackage{tabularx}
\usepackage{setspace}
\usepackage{verbatim}
\usepackage{stfloats}
\usepackage{caption}
\usepackage{float}
\usepackage[T1]{fontenc}
\usepackage{soul}
\usepackage{cite}

\usepackage{stfloats}
\usepackage{mathtools}
\usepackage{amsmath}
\allowdisplaybreaks[0]
\usepackage{subfigure}
\usepackage{fancyhdr}
\usepackage{cancel}

\makeatletter
\newif\if@restonecol
\makeatother

\usepackage[linesnumbered,ruled,vlined]{algorithm2e}
\usepackage{algpseudocode}
\usepackage{amsmath}

\usepackage{lineno}

\begin{document}

\title{Deep Reinforcement Learning Based Power Allocation for Minimizing AoI and Energy Consumption in MIMO-NOMA IoT Systems}

\author{Hongbiao Zhu, Qiong Wu, ~\IEEEmembership{Member,~IEEE}, Qiang Fan, Pingyi Fan, ~\IEEEmembership{Senior Member,~IEEE},\\ Jiangzhou Wang, ~\IEEEmembership{Fellow,~IEEE} and Zhengquan Li, ~\IEEEmembership{Member,~IEEE}
\thanks{
This work was supported in part by the National Natural Science Foundation of China under Grant No. 61701197, in part by the open research fund of State Key Laboratory of Interatory Services Networks under Grant No. ISN23-11, in part by the 111 project under Grant No. B12018, in part by the Future Network Scientific Research Fund Project (FNSRFP-2021-YB-11).

Hongbiao Zhu and Qiong Wu are with the School of Internet of Things Engineering, Jiangnan University, Wuxi 214122, China, and also with the State Key Laboratory of Integrated Services Networks (Xidian University), Xi'an 710071, China (e-mail: hongbiaozhu@stu.jiangnan.edu.cn, qiongwu@jiangnan.edu.cn).

Qiang Fan is with Qualcomm, San Jose, CA 95110, USA (e-mail: qf9898@gmail.com).

Pingyi Fan is with the Department of Electronic Engineering, Beijing National Research Center for Information Science and Technology, Tsinghua University, Beijing 100084, China (Email: fpy@tsinghua.edu.cn).

Jiangzhou Wang is with the School of Engineering, University of Kent, CT2 7NT Canterbury, U.K. (Email: j.z.wang@kent.ac.uk).

Zhengquan Li is with the School of Internet of Things Engineering, Jiangnan University, Wuxi 214122, China, and also with
Jiangsu Future Networks Innovation Institute, Nanjing 211111, China (Email: lzq722@jiangnan.edu.cn).

}
}

\markboth{IEEE XXX, ~Vol.~XX, No.~XX, XXX~2021}
{ZHU \MakeLowercase{\textit{et al.}}: Deep Reinforcement Learning Based Power Allocation for Minimizing AoI and Energy Consumption in MIMO-NOMA IoT System}


\maketitle

\begin{abstract}
Multi-input multi-out and non-orthogonal multiple access (MIMO-NOMA) internet-of-things (IoT) systems can improve channel capacity and spectrum efficiency distinctly to support the real-time applications. Age of information (AoI) is an important metric for real-time application, but there is no literature have minimized AoI of the MIMO-NOMA IoT system, which motivates us to conduct this work. In MIMO-NOMA IoT system, the base station (BS) determines the sample collection requirements and allocates the transmission power for each IoT device. Each device determines whether to sample data according to the sample collection requirements and adopts the allocated power to transmit the sampled data to the BS over MIMO-NOMA channel. Afterwards, the BS employs successive interference cancelation (SIC) technique to decode the signal of the data transmitted by each device. The sample collection requirements and power allocation would affect AoI and energy consumption of the system. It is critical to determine the optimal policy including sample collection requirements and power allocation to minimize the AoI and energy consumption of MIMO-NOMA IoT system, where the transmission rate is not a constant in the SIC process and the noise is stochastic in the MIMO-NOMA channel. In this paper, we propose the optimal power allocation to minimize the AoI and energy consumption of MIMO-NOMA IoT system based on deep reinforcement learning (DRL). Extensive simulations are carried out to demonstrate the superiority of the optimal power allocation.
\end{abstract}

\begin{IEEEkeywords}
 deep reinforcement learning, age of information, MIMO-NOMA, internet of things
\end{IEEEkeywords}

\section{Introduction}
\label{sec1}
\IEEEPARstart{W}{ith} the development of Internet-of-Things (IoT)\cite{electronics8020148,1207121,5601354}, the base station (BS) can support the real-time applications such as disaster management, information recommendation, vehicle network, smart city, connected health and industrial internet through collecting the data sampled by IoT devices\cite{Liu2019,9296803,Pingyi2018,9739659,Wu2014PerformanceMA}. However, the amount of sampled data is enormous and the number of IoT devices is usually numerous, thus the realization of these IoT applications need large bandwidth and spectrum access.  The multi-input multi-out and non-orthogonal multiple access (MIMO-NOMA) IoT can transmit data through the MIMO-NOMA channel to solve these problems, where multiple antennas are deployed at the BS to improve the channel capacity and multiple IoT devices access the common bandwidth simultaneously to improve the spectrum efficiency\cite{Zhu2021,8081737}.


The BS collects data during discrete slots in the MIMO-NOMA IoT system. In each slot, a BS first determines the sample collection requirements and allocates the transmission power for each IoT device and then sends the corresponding sample collection requirements and transmission power to each IoT device. Afterwards, each IoT device determines whether to sample data from physical world according to its sample collection requirements. Then each IoT device would adopt its allocated power to transmit the sampled data to the BS over the MIMO-NOMA channel. In the transmission process, multiple IoT devices transmit the signals of data by using a common bandwidth and thus a signal of an IoT device will be interfered by the signals of other devices. To eliminate the interference, the BS adopts the successive interference cancelation (SIC) technique to decode the received signal transmitted by each device\cite{SIC}. Specifically, the BS sorts the power of all received signals in descending order and decodes the signal with the highest received power by considering  other signals as interferences. Then the BS removes the decoded signal from the received signals and resorts the received signals to decode the next signal. The process is repeated until all signals are decoded.

The age of information (AoI) is a metric to measure the freshness of data, which is defined as the time from the data  sampling to the time when the sampled data are received. In the MIMO-NOMA IoT system, the BS needs to receive data, i.e., decode the signals of data, timely after they are sampled to provide the real-time applications, thus the MIMO-NOMA IoT system should keep low AoI \cite{AoI}\cite{Wan2022}. Furthermore, the IoT devices are energy-limited, thus the MIMO-NOMA IoT system should also keep low energy consumption to prolong the working time of IoT devices\cite{Zhu2021}. Hence, the AoI and energy consumption are two important performance metrics of the MIMO-NOMA IoT system. The sample collection requirements and power allocation would affect the AoI and energy consumption of the system\cite{8606068}. Specifically, for the sample collection requirements, if the BS selects more IoT devices to sample, the system will consume more energy because more IoT devices consume energy to sample data. However, if the BS selects less IoT devices to sample,   the data transmitted from the unselected IoT devices become obsolete, which would increase the AoI of the system. Hence the sample collection requirements affects both AoI and energy consumption of the MIMO-NOMA IoT system. For the power allocation, if an IoT device transmits with high power, the signal transmitted by the IoT device will be decoded where lots of signals with lower power act the interferences in the SIC process, which would lead to a low signal-to-interference-plus-noise ratio (SINR). Otherwise, if an IoT device transmits data with low power, the SINR would also be deteriorated due to the low transmission power. The low SINR causes a low transmission rate, which would cause a long transmission delay and  a high AoI of the MIMO-NOMA IoT system. Hence the power allocation affects the AoI of the MIMO-NOMA IoT system. Moreover, the power allocation affects the energy consumption directly. Thus, the transmission power affects both the AoI and energy consumption of the MIMO-NOMA IoT system. As mentioned above, it is critical to determine the optimal policy including sample collection requirements and power allocation to minimize the AoI and energy consumption of the MIMO-NOMA IoT system. To the best of our knowledge, there is no work to minimize the AoI in the MIMO-NOMA IoT system, which motivates us to conduct this work.

In the MIMO-NOMA IoT system, the different allocated transmission powers will impact transmission rate in the SIC process. Moreover, there is the inevitable stochastic noise in the MIMO-NOMA channel. Deep reinforcement learning (DRL) can learn the near optimal policy through considering the sample collection requirements and power allocation as action to interact with the dynamic stochastic MIMO-NOMA IoT system\cite{DRL}. In general, the DRL algorithm is suitable to solve the problem with the continuous or discrete action space exclusively. However, the space of the sample collection requirements is discrete while the space of the transmission power is continuous, which poses a challenge to the problem by DRL. In this paper, we present the relationship between the sample collection requirements and transmission power, and propose a DRL based optimal power allocation to minimize the AoI and energy consumption of the MIMO-NOMA IoT system. \footnote{The source code has been released at: https://github.com/qiongwu86/MIMO-NOMA\_AoI\_GA.git}The main contributions are summarized as follows.
\begin{itemize}
\item[1)] We formulate the joint optimization problem to minimize the AoI and energy consumption of the MIMO-NOMA IoT system by determining the sample selection and power allocation.
\item[2)] 
In the formulated  optimization problem, the sample selection is discrete and power allocation is  continuous, which can not be solve by traditional DRL method. We substitute energy model and AoI model into optimization problem, merge the  homogeneous terms containing  sample selection, and simplify the formulated problem to make it suitable to be solved by deep deterministic policy gradient (DDPG).
\item[3)] We design the DRL framework including the state, action and reward function, then adopt the DDPG algorithm to obtain the optimal power allocation to minimize the AoI and energy consumption of the MIMO-NOMA IoT system.
\item[4)] Extensive simulations are carried out to demonstrate the superiority of the optimal power allocation by the DDPG algorithm to the power allocation by the baseline algorithm.
\end{itemize}

The rest of this paper is organized as follows. Section II reviews the related work. Section III introduces the system model and formulates the optimization problem. Section IV simplifies the formulated optimization problem and presents the near optimal solution by DRL. We carry out some simulation to demonstrate the effectiveness of our proposed DRL method in Section VI, and conclude this paper in Section VII.

\section{Related Work}
In this section, we first review the studies about the AoI in the IoT system, and then survey the state of the arts on the MIMO-NOMA IoT system.

\subsection{AoI in IoT}
In \cite{generate-at-will}, Grybosi \emph{et al.} proposed the SIC-aided age-independent random access (AIRA-SIC) scheme (i.e., a slotted ALOHA fashion) for IoT system, where the receiver operates SIC to reconstruct collisions of various devices.
In \cite{wang2021}, Wang \emph{et al.} focused on the problem that minimizes the weighted sum of AoI cost and energy consumption in the IoT systems by adjusting sample policy, and proposed a distributed DRL algorithm based on the local observation of each device.
In \cite{Elmagid2020}, Elmagid \emph{et al.} aimed to minimize the AoI at the BS and the energy consumption of generate status at the IoT devices, and formulated an optimization problem based on the Markov decision process (MDP), then proved the monotonicity property of the value function associated with the MDP.
In \cite{Li2021}, Li \emph{et al.} designed a resource block (RB) allocation, modulation selecting and coding selecting scheme for each IoT device based on its channel condition to minimize the long-term AoI of IoT system.
In \cite{Hatami2021}, Hatami \emph{et al.} employed the reinforcement learning to minimize the average AoI for users in an IoT system consisting of users, energy harvesting sensors, and a cache-enabled edge node.
In \cite{Sun2021}, Sun \emph{et al.} aimed to minimize the weighted sum of the expected average AoI of all IoT devices, propulsion energy of unmanned aerial vehicle (UAV) and transmission energy of IoT devices by determining the UAV flight speed, UAV placement and channel resource allocation in the UAV-assisted IoT system.
In \cite{Hu2021}, Hu \emph{et al.} considered an IoT system where the UAVs take off from a data center to deliver energy and collect data from sensor nodes, and then fly back to the data center. They minimized the AoI of the collected data by dynamic programming (DP) and ant colony (AC) heuristic algorithms.
In \cite{Emara2020}, Emara \emph{et al.} developed a spatiotemporal framework to evaluate the peak AoI (PAoI) of the IoT system, and compared the PAoI under the time-triggered traffic with event-triggered traffic.
In \cite{Lyu2020}, Lyu \emph{et al.} considered a marine IoT scenario, where AoI is utilized to represent the impact of of the packet loss and transmission delay. They investigated the relationship between AoI and state estimation error, and minimized the state estimation error by decomposition method.
In \cite{Wang2020}, Wang \emph{et al.} investigated the impact of AoI on the system cost which consists of control cost and communication energy consumption of the industrial-internet-of-things (IIoT) system. They proved that the upper bound of cost is affected by AoI.
In \cite{Hao2022}, Hao \emph{et al.} maximized the sum of the energy efficiency of the IoT devices under the constraints of AoI by optimizing transmission power and channel allocation in a cognitive radio based IoT system.
However, none of these works have taken the MIMO-NOMA channel into account.

\subsection{MIMO-NOMA IoT System}
In \cite{Yilmaz2021}, Yilmaz \emph{et al.} proposed a user selection algorithm for MIMO-NOMA IoT system to improve the sum data rate, and adopted the physical layer network coding (PNC) to improve the spectral efficiency.
In \cite{Shi2020}, Shi \emph{et al.} considered the downlink of the MIMO-NOMA IoT networks and studied the outage probability and goodput of the system with the Kronecker model.
In \cite{wang2021_3}, Wang \emph{et al.} proposed that the resource allocation problem consists of the optimal beamforming strategy and power allocation in the MIMO-NOMA IoT system, where the beamforming optimization is solved by the zero-forcing method, after that the power allocation is solved by the convex functions.
In \cite{Han2020}, Han \emph{et al.} proposed a novel milimeter wave (mmWave) positioning MIMO-NOMA IoT system and proposed the position error bound (PEB) as a novel performance evaluation metric.
In \cite{zhang2021}, Zhang \emph{et al.}  considered the massive MIMO and NOMA to study the performance of the IoT system, and calculated the closed form function for spectral and energy efficiencies.
In \cite{Chinnadurai2018}, Chinnadurai \emph{et al.} considered the heterogeneous cellular network and formulated a problem to maximize the energy efficiency of the MIMO-NOMA IoT system, where the non-convex problem was solved based on the branch and reduced bound (BRB) approach.
In \cite{Gao2021}, Gao \emph{et al.} considered the mmWave massive MIMO and NOMA IoT system to maximize the weighted sum transmission rate through optimizing the power allocation, then solved the problem by convex method.
In \cite{Feng2021}, Feng \emph{et al.} considered an UAV aided MIMO-NOMA IoT system and regarded an UAV as BS. They formulated the problem to maximize the sum transmission rate of the downlink through optimizing the placement of UAVs, beam pattern and transmission power, then solved the problem by convex methods.
In \cite{Ding2016}, Ding \emph{et al.} designed a novel MIMO-NOMA system consisting of two different users, where user one should be served with strict quality-of-service (QoS) requirement, and user two accesses channel by the no-orthogonal way opportunistically, thus the requirement that small packets of user one in the IoT system should be transmitted in time can be met.
However, these works have not considered the AoI of the MIMO-NOMA IoT system.

As mentioned above, there is no work considering the AoI in the MIMO-NOMA IoT system, which motivates us to conduct this work.

\section{System Model And Problem Formulation}
\begin{figure}
\centering
\includegraphics[scale=0.75]{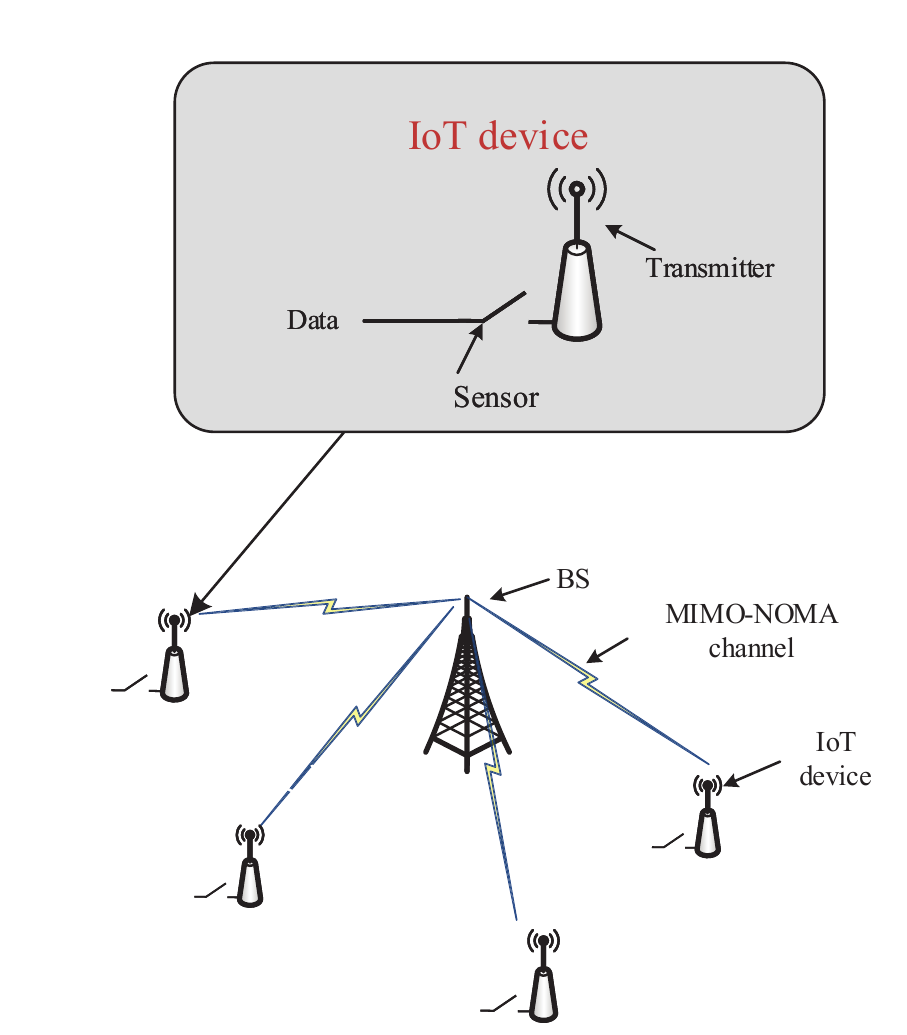}
\caption{MIMO-NOMA IoT system.}
\label{fig1}
\end{figure}

\begin{table}
\footnotesize
\caption{The summary for notations.}
\label{tab1}
\centering
\begin{tabular}{|c|p{13cm}|}
\hline
\textbf{Notation} &\textbf{Description}\\
\hline
$B$ &Population size of genetic algorithm.\\
\hline
$C_s$ &The energy consumption for sample fresh information and generate upload packet.\\
\hline
$c_{m,t}$ &Complex data symbol with 1 as variance.\\
\hline
$d_m$ &The communication distance between device $m$ and BS.\\
\hline
$E$ &Number of episodes.\\
\hline
$F_c / F_m$ &Probability of offsprings in genetic algoritm for crossover / mutation.\\
\hline
$G_P / U_P$ &Complexity of the primary networks for computing gradients / updating parameters. \\
\hline
$\boldsymbol{h}_m(t)$ &The channel vector between device $m$ and BS in slot $t$.\\
\hline
$i$ &Index of transition tuples in mini-batch.\\
\hline
$I$ &The number of transition tuples in a mini-batch.\\
\hline
$\mathcal{I}_m$ &The set of devices of which the received power is weaker than device $m$.\\
\hline
$J(\mu)$ &The long-term discounted reward under policy $\mu$.\\
\hline
$K$ &The number of antennas equipped in BS.\\
\hline
$l_{m,t}$ &The transmission delay of device $m$ in slot $t$.\\
\hline
$L$ &Loss function.\\
\hline
$l_{m,t}$ &The transmission delay of device $m$ in slot $t$.\\
\hline
$\boldsymbol{n}(t)$ &Additive white Gaussian noise.\\
\hline
$N_{GA}$ &Evolution times of genetic algorithm.\\
\hline
$m / M / \mathcal{M}$ &Index / number / set of devices.\\
\hline
$\boldsymbol{o}_t / \boldsymbol{o}_{m,t}$ &State in slot $t$ of  all devices / device $m$.\\
\hline
$\boldsymbol{p}_{t}$ / $p_{m,t}$ &Transmission power of all devices / device $m$.\\
\hline
$P_{m,t}$ &Maximum transmission power device $m$.\\
\hline
$Q(\boldsymbol{o}_t,\boldsymbol{p}_{t})$ &Action-value function under $\boldsymbol{o}_t$ and $\boldsymbol{p}_{t}$.\\
\hline
$Q$ &Packet size.\\
\hline
$r_t$ &Reward function.\\
\hline
$\boldsymbol{s}_t$ / $s_{m,t}$ &Indicator of sample or not for all devices / device $m$.\\
\hline
$S_d$ &Complexity of calculating sample decisions based on power allocation.\\
\hline
$t$ / $\mathcal{T}$ &Index / set of slot.\\
\hline
$\mathcal{U}$  &The set of undecoded received power of BS.\\
\hline
$\boldsymbol{u}_t$ / $u_{m,t}$ &Indicator of transmission success for all devices / device $m$.\\
\hline
$W$ &Bandwidth of system.\\
\hline
$\alpha_a$ / $\alpha_c$ &Learning rate of actor network / critic network.\\
\hline
$\beta$ &Discouting factor.\\
\hline
$\gamma_a$, $\gamma_c$ &Weighted factors of reward function.\\
\hline
$\Gamma_{m,t}$ &Received power of BS for device $m$ in slot $t$.\\
\hline
$\Delta_t$ &Exploration noise.\\
\hline
$\varepsilon_{m,t}$ &The energy consumed by device $m$ in slot $t$. \\
\hline
$\overline\varepsilon$ &The average sum energy consumption in slot $t$.\\
\hline
$\zeta$ / $\zeta'$ &Parameters of critic-network / target critic-network.\\
\hline
$\theta$ / $\theta'$ / $\theta^*$ &Parameters of actor-network / target actor-network /  optimal policy.\\
\hline
$\kappa$ &The constant for the update of target networks. \\
\hline
$\mu_\theta$ &Policy approximated by actor-network with $\theta$.\\
\hline
$\pi_{m,t}$ &Transmission rate of device $m$ in slot $t$.\\
\hline
$\rho_m$ &Normalized channel correlation coefficient.\\
\hline
$\sigma_{R}^{2}$ &Variance for the noise of received signal.\\
\hline
$\phi_{m,t}$ / $\Phi_{m,t}$ &AoI of device $m$ in slot $t$ on device / BS.\\
\hline
$\overline{\Phi}$ &The average sum AoI.\\
\hline
\end{tabular}
\end{table}
 
\subsection{Scenario description}
The network scenario is illustrated in Fig. \ref{fig1}. We consider a MIMO-NOMA IoT system consisting of a BS with $K$ antennas and a set $\mathcal{M}=\{1, \cdots, m, \cdots, M\}$ of the single-antenna IoT devices. Here, each IoT device is embedded with a sensor and a transmitter. The time duration is devided into $T$  slots,  each of which is $\tau$. The set of slots is denoted as $\mathcal{T}=\{1, \cdots, t, \cdots, T\}$.  At the beginning of each slot $t$, the BS determines the policy (including the sample collection requirements of each device $m$, denoted as $s_{m,t}$, and transmission power of each device $m$, denoted as $p_{m,t}$) and then sends $s_{m,t}$ and $p_{m,t}$ to each device $m$. If $s_{m,t}=1$, devic{}e $m$ will sample data in slot $t$, and transmit the data to the BS with transmission power $p_{m,t}$ over the MIMO-NOMA channel. Otherwise, it does not sample data in slot $t$. The key notations are listed in Table \ref{tab1}. Next we will construct the MIMO-NOMA channel model.
\subsection{MIMO-NOMA channel model}
Let $c_{m,t}$ be the  data symbol of device $m$ in slot $t$ with $1$ as variance, thus the signal of the data transmitted by device $m$ is $\sqrt{p_{m,t}}c_{m,t}$. Let $\boldsymbol{h}_m(t) \in \mathbb{C}^{K\times 1}$ be the channel power gain between BS and device $m$ in slot $t$, thus the corresponding signal received by BS is $\boldsymbol{h}_m(t)\sqrt{p_{m,t}}c_{m,t}$. Note that $c_{m,t}$ is unknown for BS, so that it is difficult for the BS  to calculate the received signal. Hence, the BS needs to adopt the SIC technology to decode the received signal transmitted by each device, which is expressed as
\begin{equation}
\begin{aligned}
\boldsymbol{y}(t)=\sum_{m\in\mathcal{M}}\boldsymbol{h}_m(t)\sqrt{p_{m,t}}c_{m,t}+\boldsymbol{n}(t) \\ p_{m,t} \in {[0,P_{m,max}]}, \forall m \in \mathcal{M}, \forall t \in \mathcal{T}
\label{eq1}
\end{aligned}  \quad,
\end{equation}
where $\boldsymbol{n}(t)\in \mathbb{C}^{K\times 1}$ is the complex additive white Gaussian noise (AWGN) with variance $\sigma_{R}^{2}$ and $P_{m,max}$ is the maximum transmission power of device $m$.

Similar to \cite{CSI}, it is assumed that $\boldsymbol{h}_{m}(t)$ is known by the BS. In addition, the BS also knows $p_{m,t}$, thus the BS can calculate the power of the received signal transmitted by device $m$ as
\begin{equation}
\Gamma_{m,t}=p_{m,t}||\boldsymbol{h}_{m}(t)||^2.
\label{eq4}
\end{equation}

Then, the BS decodes the received signal transmitted by each device sequentially with SIC mode of NOMA. For one iteration, the BS decodes the signal with the highest received power from $\boldsymbol{y}(t)$ while considering the other signals as interference, then removes the decoded signal from $\boldsymbol{y}(t)$ and starts the next iteration until all signals are decoded.

For instance, in an iteration the received power of the signal transmitted by device $m$ is the highest among the signals without being decoded. Denote $\mathcal{I}_m=\{k\in\mathcal{M} \mid \Gamma_{k,t}<\Gamma_{m,t} \}$ as the set of  devices whose signals' received powers is less than device $m$. Thus the signal transmitted by each device $k\in\mathcal{I}_m$ is deemed as the interference. In this case, $\boldsymbol{y}(t)$ is rewritten as
\begin{equation}
\begin{aligned}
\boldsymbol{y}(t) = \boldsymbol{h}_m(t)\sqrt{p_{m,t}}c_{m,t}+\sum_{k\in\mathcal{I}_m}\boldsymbol{h}_k(t)\sqrt{p_{k,t}}c_{k,t}+\boldsymbol{n}(t),
\label{received_power}
\end{aligned}
\end{equation}
where $\sum_{k\in\mathcal{I}_m}\boldsymbol{h}_k(t)\sqrt{p_{k,t}}c_{k,t}$ indicates the interference, thus the signal-to-interference-plus-noise ratio (SINR) of device $m$ is calculated as
\begin{equation}
\begin{split}
\gamma_{m,t}
&= \frac{p_{m,t}||\boldsymbol{h}_{m}(t)||^2}{\sum\limits_{k\in \mathcal{I}_m}p_{k,t}||\boldsymbol{h}_{k}(t)||^2+\sigma_R^2}\\
&=\frac{\Gamma_{m,t}}{\sum\limits_{k\in \mathcal{I}_m}\Gamma_{k,t}+\sigma_R^2}\\
\end{split}\quad.
\label{eq5}
\end{equation}

The transmission rate of device $m$ in slot $t$ can be derived according to Shannon capacity formula, i.e.,
\begin{equation}
\pi_{m,t} = W\log_{2}(1+\gamma_{m,t}),
\label{eq6}
\end{equation}
where $W$ is the bandwidth of the MIMO-NOMA channel.

\subsection{AoI model}
Denote $\phi_{m,t}$ as the AoI at device $m$ in slot $t$, which can be calculated as
\begin{equation}
\phi_{m,t}=
\left\{
\begin{array}{l}
{0, \qquad \qquad \qquad  s_{m,t}=1}\\
\phi_{m,t-1}+\tau, \qquad otherwise
\end{array}.
\right.\\
\label{eq7}
\end{equation}
According to Eq. \eqref{eq7}, at the beginning of slot $t$, if device $m$ samples data, i.e., $s_{m,t}=1$, $\phi_{m,t}$ will be reset to $0$. Otherwise, $\phi_{m,t}$ will be increased by $\tau$.

Device $m$ will transmit data with transmission power $p_{m,t}$ after sampling data. If data volume transmitted within a slot is larger than the packet size $Q$, i.e., $\pi_{m,t}\cdot\tau \geq Q$, device $m$ will transmit data successfully;  otherwise, the transmission is failed. Denote $u_{m,t}=1$ as a successful transmission by device $m$ in slot $t$ and $u_{m,t}=0$ as an unsuccessful transmission, we have
\begin{equation}
u_{m,t}=\left\{
\begin{array}{l}
{1, \qquad  \pi_{m,t}\cdot\tau \geq Q }\\
{0, \qquad otherwise}
\end{array}.
\right.\\
\label{eq8}
\end{equation}

If a transmission from device $m$ is successful, the AoI at the BS equals the aggregation of AoI at device $m$ and the transmission delay. Otherwise, the AoI at the BS is increased by a slot, thus we have \begin{equation}
\Phi_{m,t}=\left\{
\begin{array}{l}
{\phi_{m,t}+ l_{m,t}, \qquad  u_{m,t}=1 }\\
{\Phi_{m,t-1}+\tau, \qquad   otherwise}
\end{array},
\right.\\
\label{eq9}
\end{equation}
where $l_{m,t}$ is the transmission delay of device $m$ in slot $t$, which is calculated as
\begin{equation}
l_{m,t} = \frac{Q}{\pi_{m,t}}.
\label{eq10}
\end{equation}

The AoI of the MIMO-NOMA IoT system is measured by averaging the AoI of all devices  at BS, i.e.,
\begin{equation}
\overline{\Phi}= \frac{1}{T}\sum_{t\in\mathcal{T}}\sum_{m\in\mathcal{M}}\Phi_{m,t}
\label{eq11}.
\end{equation}

\subsection{Energy consumption model}
Since each device consumes energy in data sampling and transmission, the energy consumption of device $m$ in slot $t$ can be calculated as
\begin{equation}
\varepsilon_{m,t} = s_{m,t}C_s + p_{m,t}l_{m,t},
\label{eq12}
\end{equation}
where $C_s$ is the energy consumption for data sampling\cite{wang2021}, and $p_{m,t}l_{m,t}$ is the energy consumption for transmission.

The BS has a stable power supply, hence the energy consumption of BS is sufficient and thus it is not taken into account in the system. Hence the energy consumption of the MIMO-NOMA IoT system is measured by averaging the energy consumption of all devices, i.e.,
\begin{equation}
\overline\varepsilon = \frac{1}{T}\sum_{t\in\mathcal{T}}\sum_{m\in\mathcal{M}}\varepsilon_{m,t}.
\label{eq13}
\end{equation}

\subsection{Problem formulation}
In this work, our target is to minimize the AoI and energy consumption of the MIMO-NOMA IoT system, which is impacted by $p_{m,t}$ and $s_{m,t}$. Therefore the optimization problem is formulated as
\begin{align}
\label{eq14}
&\min_{\boldsymbol{s}_t,\boldsymbol{p}_t} \left[\gamma_a\overline{\Phi}+\gamma_e\overline\varepsilon \right]\\
s.t. \qquad &p_{m,t}\in [0,P_{m,max}], \forall m \in \mathcal{M}, \forall t \in \mathcal{T}, \tag{\ref{eq14}{a}} \label{eq14a}\\
&s_{m,t}\in\{0,1\}, \forall m \in \mathcal{M}, \forall t \in \mathcal{T} \tag{\ref{eq14}{b}} \label{eq14b},
\end{align}
where $\boldsymbol{s}_t = \{s_{1,t}, \cdots, s_{m,t}, \cdots, s_{M,t}\}$ and $\boldsymbol{p}_t = \{p_{1,t}, \cdots, p_{m,t}, \cdots, p_{M,t}\}$, $\gamma_a$ and $\gamma_e$ are the non-negative weighted factors. Next we will present a solution to the problem based on DRL.

\section{DRL Method for Optimization of Power Allocation}
\label{sec4}
In this section, we solve the optimization problem based on the DRL. First, we design the DRL framework including the state, action and reward function, where the relationship between the sample collection requirements and transmission power is derived to facilitate DRL algorithm  to solve the problem.  Then we obtain a near optimal power allocation based on the DRL algorithm.
\subsection{DRL framework}
The DRL framework  consists of three significant elements: state, action and reward function. For each slot, the agent observes the current state and takes the current action according to policy $\mu$, where policy $\mu$ yields the  action based on the  state. Then the agent calculates its corresponding reward under the current state and action according to the reward function, while the current state in the environment transits to the next state. Next, we will design agent, state, action and reward function, respectively.
\begin{itemize}
\item \textbf{Agent:} In each slot, the BS determines the transmission power and sample collection requirements of each device based on its observation, thus we consider the BS as the agent.
\item \textbf{State:} In the system model, the state $\boldsymbol{o}_t$ observed by the BS in slot $t$ is defined as
\begin{equation}
\boldsymbol{o}_t = [\boldsymbol{o}_{1,t}, \cdots, \boldsymbol{o}_{m,t}, \cdots, \boldsymbol{o}_{M,t}],
\label{eq15}
\end{equation}
where $\boldsymbol{o}_{m,t}$ represents the observation of device $m$, which is designed as
\begin{equation}
\boldsymbol{o}_{m,t} = [u_{m,t-1}, \gamma_{m,t-1}, \Phi_{m,t-1}].
\label{eq16}
\end{equation}
Here, $u_{m,t-1}$, $\gamma_{m,t-1}$ and $\Phi_{m,t-1}$ can be calculated by the BS from the historical data in slot $t-1$.
\item \textbf{Action:} According to the problem formulated in Eq. \eqref{eq14}, the action in slot $t$ is set as
\begin{equation}
\boldsymbol{a}_t = [\boldsymbol{s}_t, \boldsymbol{p}_t].
\label{action}
\end{equation}
\begin{figure*}[ht]
\begin{align}
\label{proof1}
&\gamma_a\overline{\Phi}+\gamma_e\overline\varepsilon\\
=&\frac{1}{T}\sum_{t\in\mathcal{T}}\sum_{m\in\mathcal{M}}\Big[\gamma_a\Phi_{m,t} + \gamma_e\varepsilon_{m,t}\Big] \tag{\ref{proof1}{a}} \label{proof1a}\\
=&\frac{1}{T}\sum_{t\in\mathcal{T}}\sum_{m\in\mathcal{M}}\Big[\gamma_a\big[(1-u_{m,t})(\Phi_{m,t-1}+\tau)+u_{m,t}(\phi_{m,t}(s_{m,t})+ l_{m,t})\big] + \gamma_e(s_{m,t}C_s + p_{m,t}l_{m,t})\Big] \tag{\ref{proof1}{b}} \label{proof1b}\\
=&\frac{1}{T}\sum_{t\in\mathcal{T}}\sum_{m\in\mathcal{M}}\Big[[\gamma_a u_{m,t} \phi_{m,t}(s_{m,t}) +\gamma_e s_{m,t}C_s] + \gamma_a[(1-u_{m,t})(\Phi_{m,t-1}+\tau)+u_{m,t}l_{m,t}] +\gamma_ep_{m,t}l_{m,t}\Big] \tag{\ref{proof1}{c}} \label{proof1c}
\end{align}
\end{figure*}
\begin{figure*}
\begin{align}
\label{proof2}
&\gamma_a\Phi_{m,t}(s_{m,t},p_{m,t}) + \gamma_e\varepsilon_{m,t}(s_{m,t}, p_{m,t})\\
=&\gamma_au_{m,t}(1-s_{m,t})(\phi_{m,t-1}+\tau)+\gamma_es_{m,t}C_s+ \gamma_a[(1-u_{m,t})(\Phi_{m,t-1}+\tau)+u_{m,t}l_{m,t}] +\gamma_ep_{m,t}l_{m,t} \tag{\ref{proof2}{a}} \label{proof2a}\\
=&s_{m,t}[\gamma_eC_s-\gamma_au_{m,t}(\phi_{m,t-1}+\tau)]+\gamma_a[u_{m,t}(\phi_{m,t-1}+\tau)+(1-u_{m,t})(\Phi_{m,t-1}+\tau)+u_{m,t}l_{m,t}] +\gamma_ep_{m,t}l_{m,t} \tag{\ref{proof2}{b}} \label{proof2b}\\
=&s_{m,t}C_{m,t,1} +C_{m,t,2} \tag{\ref{proof2}{c}} \label{proof2c}
\end{align}
\rule[-10pt]{18.1cm}{0.05em}
\end{figure*}

The traditional two DRL algorithms DDPG and Deep Q-Learning (DQN) are suitable for continuous and discrete action space, respectively. However, $s_{m,t}\in\{0,1\}$ and $p_{m,t}\in[0,P_{m,max}]$ in Eq. \eqref{action}, thus the space of $\boldsymbol{s}_t$ is discrete while the space of $\boldsymbol{p}_t$ is continuous. Hence, the optimization problem can neither be solved by DQN nor DDPG. Next, we will investigate the relationship between $p_{m,t}$ and $s_{m,t}$ to handle this dilemma.

Substituting Eqs. \eqref{eq11} and \eqref{eq13} into Eq. \eqref{eq14}, the optimization objective is  rewritten as Eq. \eqref{proof1a}. Then substituting Eqs. \eqref{eq9} and  \eqref{eq12} into Eq. \eqref{proof1a}, we can obtain Eq. \eqref{proof1b}, where $\phi_{m,t}$ is denoted as $\phi_{m,t}(s_{m,t})$ to indicate that it is the function of $s_{m,t}$. Then reorganizing Eq. \eqref{proof1b}, we have Eq. \eqref{proof1c}. The first term of Eq. \eqref{proof1c} is related with $s_{m,t}$, next we rewrite the first term of  Eq. \eqref{proof1c} as Eq. \eqref{proof2} to investigate the relationship between $s_{m,t}$ and $p_{m,t}$. Substituting Eq. \eqref{eq7} into Eq. \eqref{proof2}, we have Eq. \eqref{proof2a}. Then merge the homogeneous terms containing $s_{m,t}$ and $\gamma_a$ in Eq. \eqref{proof2a}, respectively, we have Eq. \eqref{proof2b}. Let $C_{m,t,1}=\gamma_eC_s-\gamma_au_{m,t}(\phi_{m,t-1}+\tau)$ and $C_{m,t,2}=\gamma_a[u_{m,t}(\phi_{m,t-1}+\tau)+(1-u_{m,t})(\Phi_{m,t-1}+\tau)+u_{m,t}l_{m,t}] +\gamma_ep_{m,t}l_{m,t}$, thus Eq. \eqref{proof2b} is rewritten as Eq. \eqref{proof2c}, where $C_{m,t,1}$ is the coefficient for homogeneous terms containing $s_{m,t}$ in Eq. \eqref{proof2b}, and $C_{m,t,2}$ contains all terms without $s_{m,t}$ in Eq. \eqref{proof2b}.

In $C_{m,t,1}$ and $C_{m,t,2}$, $\phi_{m,t-1}$ can be calculated by the BS based on the historical data in slot $t-1$\cite{wang2021} and $\Phi_{m,t-1}$ is known for the BS. In addition, BS can calculate $\gamma_{m,t}$ according to Eqs. \eqref{eq5}-\eqref{eq6}, thus $u_{m,t}$ and $l_{m,t}$ can be further calculated according to Eqs. \eqref{eq8} and \eqref{eq10}  given ${p}_{m,t}$, which means that $C_{m,t,1}$ and $C_{m,t,2}$ depend on ${p}_{m,t}$ and are independent of $s_{m,t}$. Hence, the optimal sample collection requirements to minimize $s_{m,t}C_{m,t,1} +C_{m,t,2}$, denoted as $s^*_{m,t}$, is achieved when the term $s_{m,t}C_{m,t,1}$ is minimum, thus we have
\begin{equation}
s^*_{m,t}=\left\{
\begin{array}{l}
{1, \qquad  C_{m,t,1} < 0 }\\
{0, \qquad otherwise}
\end{array}.
\right.
\label{proof4}
\end{equation}
Hence, the optimal sample collection requirements can be determined according to Eq. \eqref{proof4} when ${p}_{m,t}$ is given and Eq. \eqref{eq14} can be rewritten as
\begin{align}
\label{eq18}
&\min_{\boldsymbol{p}_t} \left[\gamma_a\overline{\Phi}+\gamma_e\overline\varepsilon \right]\\
s.t.  \qquad &p_{m,t}\in [0,P_{m,max}], \forall m \in \mathcal{M}, \forall t \in \mathcal{T}, \tag{\ref{eq18}{a}}\\ &s^*_{m,t}=\left\{
\begin{array}{l}
{1, \qquad  C_{m,t,1} < 0 }\\
{0, \qquad otherwise}
\end{array}. \tag{\ref{eq18}{b}}
\right.
\end{align}
 According to Eq. \eqref{eq18}, the action $\boldsymbol{a}_t$ is only reflected by  $\boldsymbol{p}_t$. Therefore, DDPG which is suitable for the continuous action space can be employed as the desired algorithm to solve the optimization problem in Eq. \eqref{eq18}.

\item \textbf{Reward function:}
The BS aims to minimize the AoI and energy consumption of the MIMO-NOMA IoT system, and the target of the DDPG algorithm is to maximize the reward function. Therefore the reward function in slot $t$ can be defined as

\begin{equation}
{r}_{t}(\boldsymbol{o}_t,\boldsymbol{p}_t)=-\sum_{m\in\mathcal{M}}[\gamma_a\Phi_{m,t} + \gamma_e\varepsilon_{m,t}].
\label{eq19}
\end{equation}

Furthermore, the expected long-term discounted reward of the system can be defined as
\begin{equation}
J(\mu)= \mathbb{E}\left[\sum_{t=1}^{T}\beta^{t-1}r_{t}(\boldsymbol{o}_t,\boldsymbol{p}_t)|_{\boldsymbol{p}_t=\mu(\boldsymbol{o}_t)}\right],
\label{eq20}
\end{equation}
where $\beta\in[0,1]$ is the discounting factor, $\boldsymbol{p}_t= \mu(\boldsymbol{o}_t)$ indicates the action under the state $\boldsymbol{o}_t$, which is derived through policy $\mu$. Thus, our objective in this paper becomes finding the optimal policy to minimize $J(\mu)$.

\end{itemize}

\subsection{Optimizing power allocation based on DDPG}
\begin{figure*}
\centering
\includegraphics[width=7in]{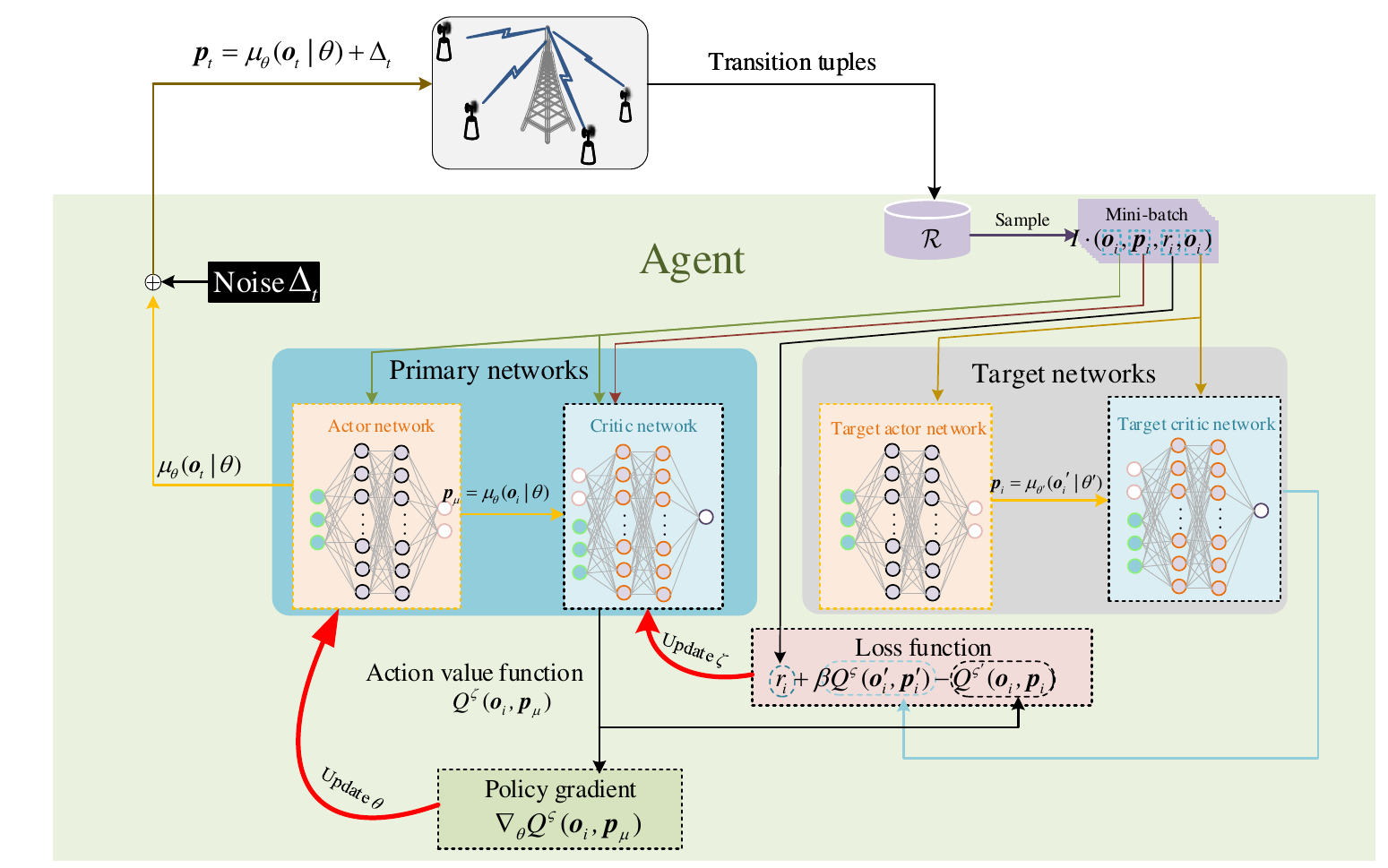}
\caption{Flow diagram of DDPG}
\label{fig2}
\end{figure*}
In this subsection, we will introduce the architecture of the DDPG algorithm including primary networks (a actor network and a critic network) and target networks (a target actor network and a target critic-network)\cite{Qiao2020}, where the actor network is adopted for policy approximation and improvement, the critic network is adopted for policy evaluation and the target networks are adopted to improve the stability of algorithm. Both primary and target networks are neural networks (DNNs). Denote $\theta$, $\zeta$, $\theta'$ and $\zeta'$ as  parameters of the actor network, critic network, target actor network and target critic network, respectively, $\mu_\theta$ as the policy approximated by actor-network, and $\Delta_t$ as the noise added on action for exploration in slot $t$. Next, we will present the training stage of the DDPG algorithm in detail.

The parameters $\theta$ and $\zeta$ are first initialized randomly, $\theta'$ and $\zeta'$ are set as $\theta$ and $\zeta$, respectively. In addition, a replay experience buffer $\mathcal{R}$ is built up to cache the state transitions (lines 1-3).

Next, the algorithm loops for $E$ episodes. At the beginning of each episode, the simulation parameters of the system model is reset as $u_{m,0} = 0$, $p_{m,0}=1$ and $\Phi_{m,0}=0$ for each device $m$, $\boldsymbol{h}_{m}(0)$ is initialized randomly. Given $p_{m,0}$ and $\boldsymbol{h}_{m}(0)$, the SINR $\gamma_{m,0}$ is calculated according to Eqs. \eqref{eq4}-\eqref{eq5},  then the state of each device $m$, i.e., $\boldsymbol{o}_{m,1}=[u_{m,0}, \gamma_{m,0}, \Phi_{m,0}]$ is observed by the agent (lines 4-6).

Afterwards, the algorithm iterates from slot $1$ to $T$. For slot $t$, the actor network yields the output $\mu_{\theta}(\boldsymbol{o}_t|\theta)$ under the observed state $\boldsymbol{o}_t$ and policy $\mu$ with parameters $\theta$. Then a noise $\Delta_t$ is generated and the agent calculates the transmission powers of all devices according to $\boldsymbol{p}_t=\mu_{\theta}(\boldsymbol{o}_t|\theta)+\Delta_t$. After that the agent  calculates $u_{m,t}$, $s_{m,t}$ and $\gamma_{m,t}$ of each device $m$ according to Eqs. \eqref{eq8}, \eqref{proof4} and \eqref{eq5}, respectively. Afterwards, the agent calculates $\Phi_{m,t}$ and $\varepsilon_{m,t}$ according to Eqs. \eqref{eq9} and \eqref{eq12}, respectively, and thus obtain the state of slot $t$, i.e., $\boldsymbol{o}_{t+1}$, then calculates $r_t$ according to Eq. \eqref{eq19}.  The above tuple $[\boldsymbol{o}_t, \boldsymbol{p}_t, r_t, \boldsymbol{o}_{t+1}]$ in the replay buffer. Then the agent inputs $\boldsymbol{o}_{t+1}$ into the actor network and starts the next iteration if the number of samples in replay buffer is not larger than $I$ (lines 7-10).

If the number of tuples in the replay buffer exceeds $I$, the parameters $\theta$, $\theta'$, $\zeta$, and $\zeta'$ will be updated to maximize $J(\mu_\theta)$. Here, $\theta$ is updated toward the direction of the gradient $\nabla_{\theta}J(\mu_\theta)$. Specifically, the agent uniformly retrieves a mini-batch consisting of $I$ tuples from the replay buffer. For each tuple $i$, i.e., $(\boldsymbol{o}_{i}, \boldsymbol{p}_{i}, r_i, \boldsymbol{o}_{i}')$ $(i\in\{1, 2, \cdots, I\})$, the agent inputs $\boldsymbol{o'}_{i}$ into the target actor network and outputs $\boldsymbol{p}_{i}'=\mu_{\theta'}({\boldsymbol{o'}_{i}}|\theta')$, then inputs $\boldsymbol{o}_{i}'$ and $\boldsymbol{p}_{i}'$ into the target critic network and outputs  $Q^{\zeta'}({\boldsymbol{o}_{i}'},\boldsymbol{p}_{i}')$, then calculates the target value as
\begin{equation}
y_{i}=r_{i}+\beta Q^{\zeta'}({\boldsymbol{o}_{i}'},\boldsymbol{p}_{i}')|_{\boldsymbol{p}_{i}'=\mu_{\theta'}({\boldsymbol{o'}_{i}}|\theta')}.
\label{eq22}
\end{equation}

While $\boldsymbol{o}_{i}$ and $\boldsymbol{p}_{i}$ are the input and  $Q^{\zeta}(\boldsymbol{o}_{i},\boldsymbol{p}_{i})$ is the output of critic-network, the loss function can be expressed as 
\begin{equation}
L(\zeta)=\frac{1}{I}\sum_{i=1}^{I}\left[y_{i}-Q^{\zeta}(\boldsymbol{o}_{i},\boldsymbol{p}_{i})\right]^2.
\label{eq23}
\end{equation}
Then the critic network is updated by the gradient descending method with the gradient of loss function $\nabla_{\zeta}L(\zeta)$ \cite{DDPG} (lines 11-13), i.e.,
\begin{equation}
\zeta \leftarrow \zeta - \alpha_c\nabla_{\zeta}L(\zeta),
\label{loss_grad}
\end{equation}
where $\alpha_c$ is the learning rate of the critic network.


After that, the agent calculates the gradient $\nabla_{\theta}J(\mu_\theta)$ as  \cite{D.silver}
\begin{equation}
\begin{split}
&\nabla_{\theta}J(\mu_{\theta})\\
&\approx \frac{1}{I}\sum_{i=1}^{I}\nabla_{\theta}Q^{\zeta}(\boldsymbol{o}_{i},\boldsymbol{p}_{\mu})|_{\boldsymbol{p}_{\mu}=\mu_{\theta}(\boldsymbol{o}_{i}|\theta)}\\
&=\frac{1}{I}\sum_{i=1}^{I}\nabla_{\theta}\mu_{\theta}(\boldsymbol{o}_{i}|\theta) \cdot \nabla_{\boldsymbol{p}_{\mu}}Q^{\zeta}(\boldsymbol{o}_{i},\boldsymbol{p}_{\mu})|_{\boldsymbol{p}_{\mu}=\mu_{\theta}(\boldsymbol{o}_{i}|\theta)}
\end{split},
\label{eq24}
\end{equation}
where chain rule is applied to derive the gradient of $Q^{\zeta}(\boldsymbol{o}_{i},\boldsymbol{p}_{\mu})$ with respect to $\theta$ \cite{D.silver}. Given $\nabla_{\theta}J(\mu_{\theta})$, actor-network can be updated by gradient ascending to maximize $J(\mu_{\theta})$, i.e,
\begin{equation}
\theta \leftarrow \theta + \alpha_a\nabla_{\theta}J(\mu_{\theta}),
\label{obj_grad}
\end{equation}
where $\alpha_a$ is the learning rate of actor network.

After the parameters of the primary networks are updated, the parameters of the target networks are updated based on the parameters of primary networks, i.e.,
\begin{equation}
\begin{split}
\zeta'&\leftarrow \kappa\zeta + (1-\kappa)\zeta'\\
\theta^{\prime}&\leftarrow\kappa\theta+(1-\kappa)\theta^{{\prime}}
\label{eq25}
\end{split} \quad ,
\end{equation}
where $\kappa$ is a constant much smaller than 1, i.e., $\kappa \ll 1$. (line 15).

Up to now, the iteration for slot $t$ is finished and the agent starts the next iteration until the number of slots reaches $T$. Then the agent starts the next episode. When the number of episodes reaches $E$, the training stage is finished and outputs the near optimal policy.
The pseudocode of the training stage is described in Algorithm \ref{al1}.

\begin{algorithm}
  \caption{Training stage of the DDPG algorithm}
  \label{al1}
  \KwIn{$\gamma$, $\tau$, $\theta$, $\zeta$}
  \KwOut{optimized DNNs}
  Randomly initialize the $\theta$, $\zeta$\;
  Initialize target networks by $\zeta'\leftarrow\zeta$, $\theta'\leftarrow\theta$\;
  Initialize replay experience buffer $\mathcal{R}$\;
  \For{episode from $1$ to $E$ }
  {
    Reset simulation parameters for the system model\;
    Receive initial observation state $\boldsymbol{o}_{1}$\;
    \For{slot $t$ from $1$ to $T$ }
    {
      Generate the transmission power of all devices according to the current policy, state and exploration noise $\boldsymbol{p}_t=\mu_{\theta}(\boldsymbol{o}_t|\theta)+\Delta_{t}$ \;
      Execute action $\boldsymbol{p}_t$, observe reward $r_t$ and new state $\boldsymbol{o}_{t+1}$ from the system model\;
      Store transition tuple $(\boldsymbol{o}_t,\boldsymbol{p}_t,r_t,\boldsymbol{o}_{t+1})$ in $\mathcal{R}$\;
      \If {number of tuples in $\mathcal{R}$ is larger than $I$ }
      {
      Randomly sample a mini-batch of $I$ transitions tuples from $\mathcal{R}$\;
      Update the critic network by minimizing the loss function according to Eq. \eqref{loss_grad}\;
      Update the actor network according to Eq. \eqref{obj_grad}\;
      Update target networks according to Eqs. \eqref{eq25}.
      }
    }
  }
\end{algorithm}

Next, the testing stage is initialized to test the performance under the near optimal policy. Compared with the training stage, the parameter updating process is omitted in testing process and actions in each slot are generated by the near optimal policy. The corresponding pseudocode is shown in Algorithm \ref{al2}, where $\theta^*$ is the parameter to achieve the near optimal policy in the training stage.
\begin{algorithm}
  \caption{Testing stage of the DDPG algorithm}
  \label{al2}
  \For{episode from $1$ to $E$ }
  {
    Reset simulation parameters for the system model\;
    Receive initial observation state $\boldsymbol{o}_{1}$\;
    \For{ slot $t$ from $1$ to $T$ }
    {
      Generate the transmission power of all devices according to the near optimal policy and current state $\boldsymbol{p}_t=\mu_{\theta}(\boldsymbol{o}_{t}|\theta^{m*})$ \;
      Execute the action $\boldsymbol{p}_t$;\\
      Observe reward the $r_t$ and new state $\boldsymbol{o}_{t+1}$.
    }
  }
\end{algorithm}
{\subsection{Complexity Investigation}}
{In this subsection, we investigated the complexity of proposed algorithm. Denote $G_P$ and $U_p$ as the computational complexity for computing gradients and updating parameters of the primary networks, respectively. Since the architecture of target networks are same as primary networks, the computational complexity for updating parameters of the target networks are the same as primary networks. The complexity of proposed algorithm is related to the number of slots in training process. To be specific, during each slot the primary networks calculate gradients and updating parameters, while the target networks update parameters with the parameters of primary networks according to  Eq. \eqref{eq25}. Moreover, denote the complexity of calculating sample decisions based on power allocation as $S_d$. Thus the complexity of the proposed algorithm in a slot is $O(G_P + 2U_P + S_d)$. Note that the gradients calculating and parameters updating will be processed until number of tuples cached in the replay buffer exceeds I, and the proposed algorithm will loop $E$ episodes which each of episodes contains T slots, to this end the complexity of proposed algorithm can be expressed as $O((E\cdot T - I)( G_P + 2U_P + S_d))$.}

\section{Simulation Results and Analysis}
\label{sec6}
In this section, we will conduct simulation to verify the effectiveness of our proposed power allocation policy. The experiments consist of both the training and testing stage. The simulation scenario is described in the system model and the simulation tool is Python 3.6. In the simulation, both actor network and critic network are the four-layer fully connected DNN with two hidden layers which are equipped with $400$ and $300$ neurons, respectively. Adam optimization method \cite{adam} is adopted to update the parameters of critic network and actor network.
The noise $\Delta_{t}$ (for exploration) follows the Ornstein-Uhlenbeck (OU) process with decay-rate $0.15$ and variation $0.004$, respectively \cite{OUnoise}. The small scale fading of each device is initialized by white Gaussian noise, and the Rayleigh block fading model is employed to simulate the stochastic small scale fading\cite{ARmodel}. The reference channel gain of each device is $-30 dB$ when the communications distance is $1$ meter, the path-loss exponent is $2$, and the communication distances is randomly set within a range $[50,100]$ meters. The parameters are summarized in TABLE \ref{tab2}.
\begin{figure}
\centering
\includegraphics[scale=0.5]{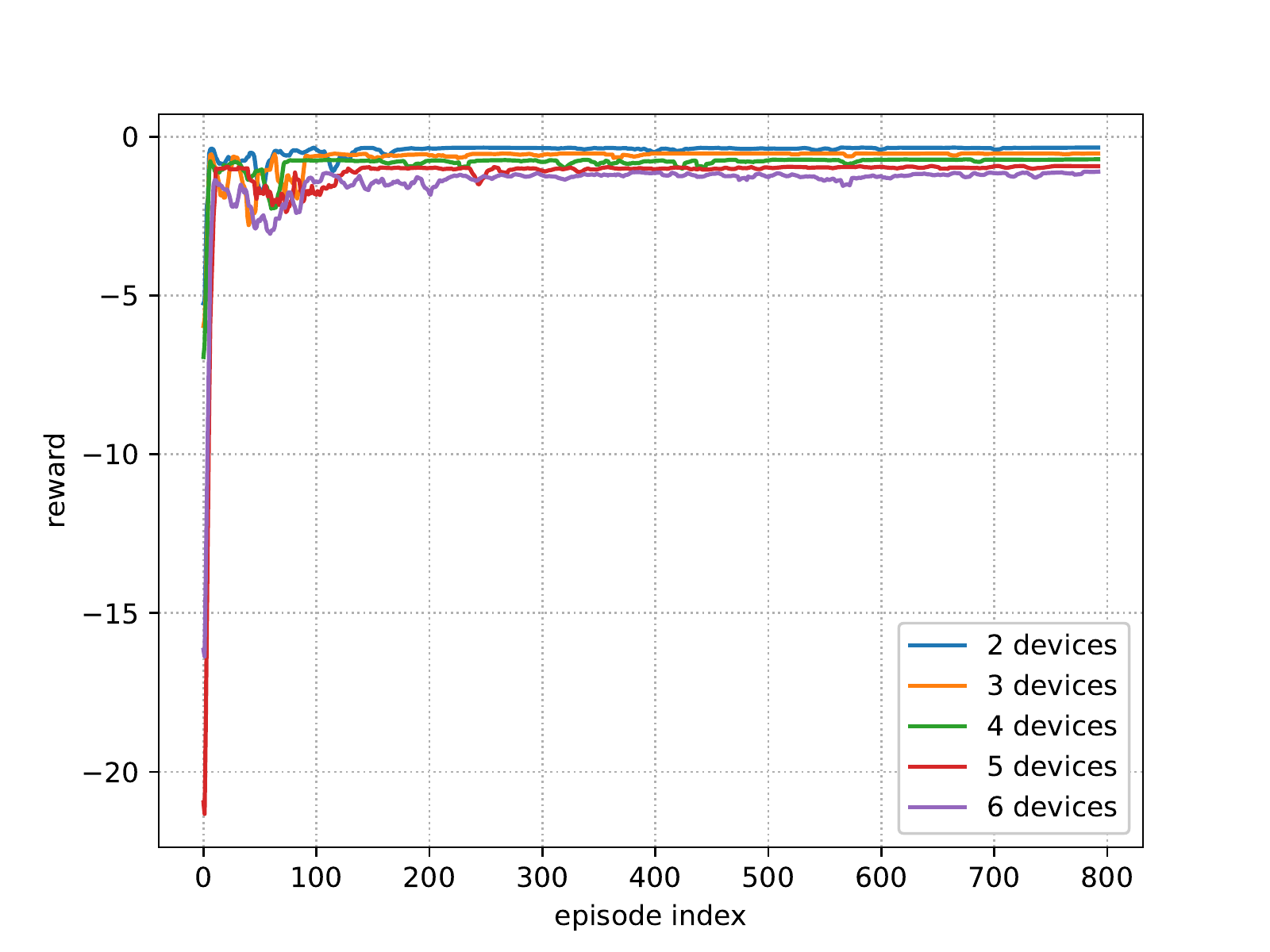}
\caption{Learning curves under various number of devices.}
\label{fig9}
\end{figure}
\begin{table}
\caption{Values of the parameters in the experiments.}
\label{tab2}
\footnotesize
\centering
\begin{tabular}{|c|c|c|c|}
  \hline
  \multicolumn{4}{|c|}{Parameters of system model{\cite{9691928}}}\\
  \hline
  \textbf{Parameter} &\textbf{Value} &\textbf{Parameter} &\textbf{Value}\\
  \hline
  $\tau$ &$0.1$ s &$K$ &$4$\\
  \hline
  $W$ &$18$ kHz &$C_s$ &0.5 J\\
  \hline
  $P_{m,max}$ &$2$ W &$T$ &$500$ \\
  \hline
  \multicolumn{4}{|c|}{Parameters of Agent{\cite{DDPG}}}\\
  \hline
  \textbf{Parameter} &\textbf{Value} &\textbf{Parameter} &\textbf{Value}\\
  \hline
  $\kappa$ &$0.001$  & $I$ &$64$ \\
  \hline
  $E$ &$800$ &$\beta$ &$0.99$\\
  \hline
  $|\mathcal{R}|$ &$2.5\times10^5$ &$\gamma_e$ &$0.5$\\
  \hline
  $\gamma_a$ &$0.5$  &$\alpha_a$ & $10^{-3}$\\
  \hline
  $\alpha_c$ &$10^{-4}$ &{$F_c / F_m$} &{$0.8 / 0.5$} \\
  \hline
  {$B$} &{$10$} &{$N_{GA}$} &{50}\\
  \hline
\end{tabular}
\end{table}

\subsection{Training Stage}
Fig. \ref{fig9} shows the learning curves in the training stage, i.e., rewards in different episode, for different numbers of IoT devices. It can be seen that the rewards of different curves rise and fluctuate from episode $0$ to $150$, which reflects that the agent is learning the policy to maximize the average reward. After that the learning curves turn out to be stable, which indicates that the near optimal policy has been learned by the agent. Note that there is a litter jitters after episode $150$, it is because the agent is adjusting slightly since the exploration noise prevents the agent from converging into the local optima. It also can be seen that the large number of devices incurs a low reward. It is attributed to the fact that each device will be affected by more interference as the number of devices in the system increases, which leads to the lower transmission rate. It will prolongs the transmission delay and further increase the AoI of the system. Then the BS would inform the devices to consume more energy to sample more frequently and transmit faster, thus the lower AoI can be guaranteed.

\subsection{Testing Stage}
In the testing stage, we check the performance of the near optimal policy obtained in the training stage. {Random power allocation policy and Genetic Algorithm (GA) \cite{RAHNAMAYAN20071605} are adopted for comparison, random power allocation policy and GA are introduced as follow:}

\begin{itemize}
\item {Random policy: Randomly allocates the power of each device $m$ within $[0,P_{m,max}]$ and the sample collection requirements is obtained according to Eq. (19).}
\item {Genetic Algorithm: In each time slot, BS randomly generates population vector according to $P_{m, max}$ and population size $B$, which means that there are $B$ individuals in the population vector, and each individual in the population vector is the power allocation of all device. BS selects best individuals in the population vector as offsprings according to fitness, i.e., reward function, of each individuals. Then offsprings will evolve for $N_{GA}$ times, for each evolution, the probabilities of crossover and mutation for these offsprings are $F_c$ and $F_m$, respectively. Where crossover means that two individuals in the offsprings exchange power allocation of a random device, and mutation means that the power allocation of any device in the offsprings sets  in $[0, P_{m,max}]$ randomly. After that, selecting best individuals from the offsprings that has experienced crossover and mutation as next offsprings and start next evolution. After all evolutions, select best individual from the last offsprings, which is the optimal power allocation derived by GA. After that, BS calculates the optimal sample collection based on the optimal power allocation according to Eq. (19), then executes the optimal power allocation derived by GA and optimal sample collection. In the end, BS iterates into the next time slot.} 
\end{itemize}

\begin{figure}
\centering
\includegraphics[scale=0.5]{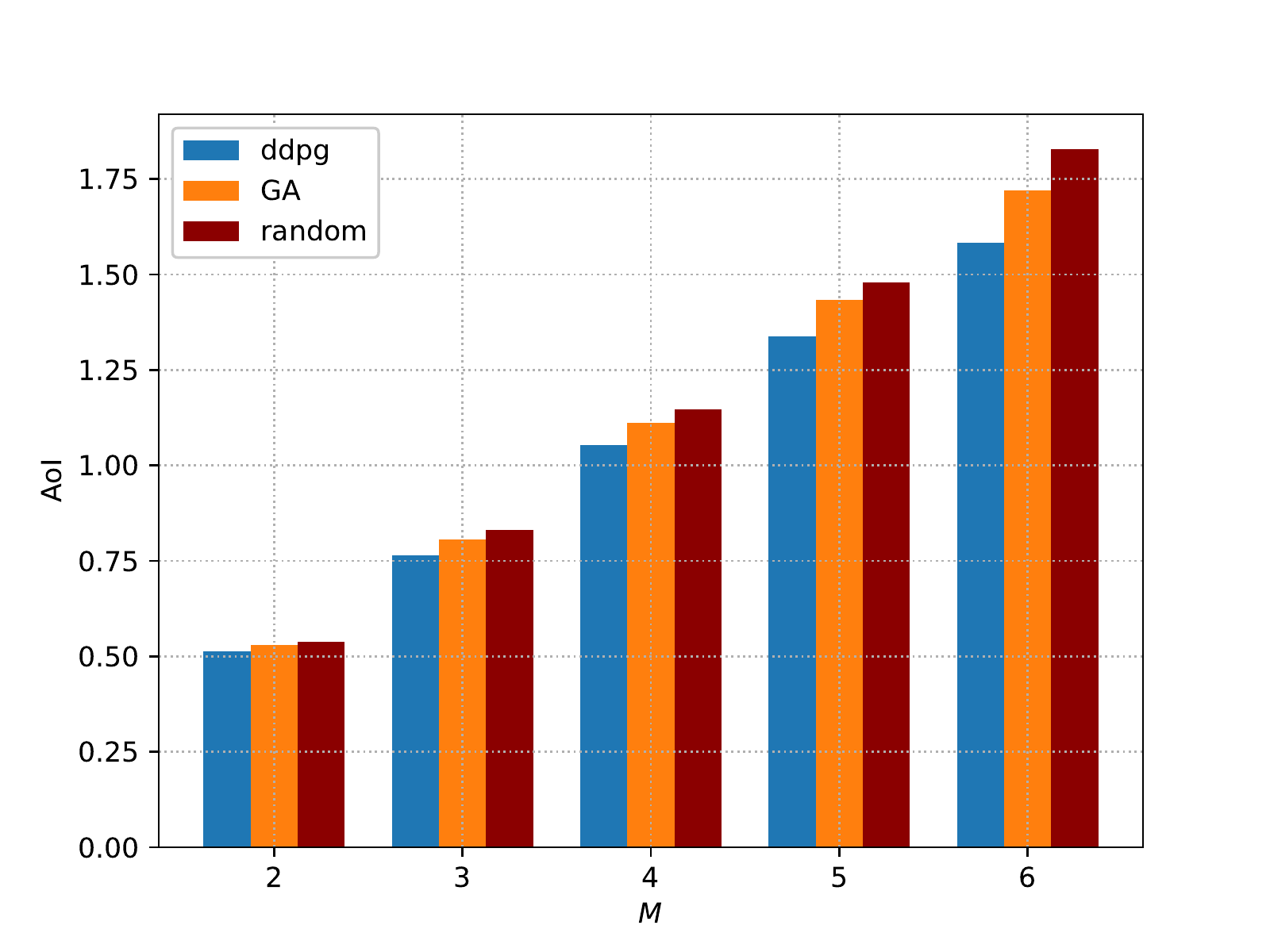}
\caption{AoI of the system vs. number of devices.}
\label{fig10}
\end{figure}

Fig. \ref{fig10} presents the AoI of near optimal policy with that of the random policy {and GA}. It can be seen that the AoI of the system under the { three} policies increases as the number of devices increases. This is because that each device will suffer from the interference as the number of devices increases, and thus degrades its transmission delay according to Eq. \eqref{eq10}, which would further increase the AoI of the system. {Meanwhile, the near optimal policy and GA always outperform the random policy, because optimal policy can adjust the power allocation adaptively according to the observed state, and GA will find the optimal power allocation according to fitness in the evolutions to ensure low AoI, while random policy just generates power allocation randomly. It also can be seen that the near optimal policy outperforms the GA, because GA may converge into local optima, while DDPG has the exploration noise to prevent the agent from converging into the local optima. }

\begin{figure}
\centering
\includegraphics[scale=0.5]{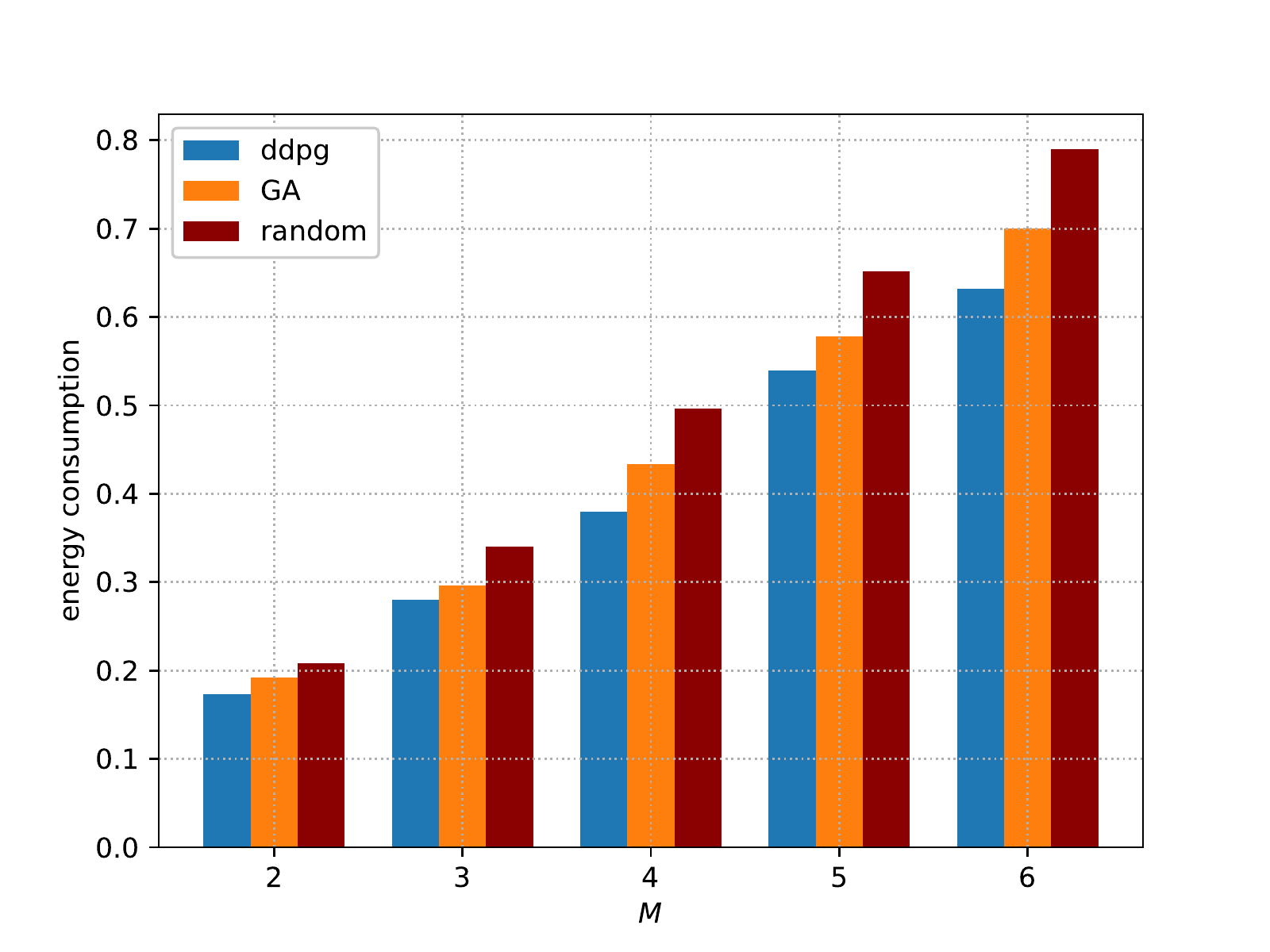}
\caption{Energy consumption of the system vs. number of devices.}
\label{fig11}
\end{figure}

\begin{figure}
\centering
\includegraphics[scale=0.5]{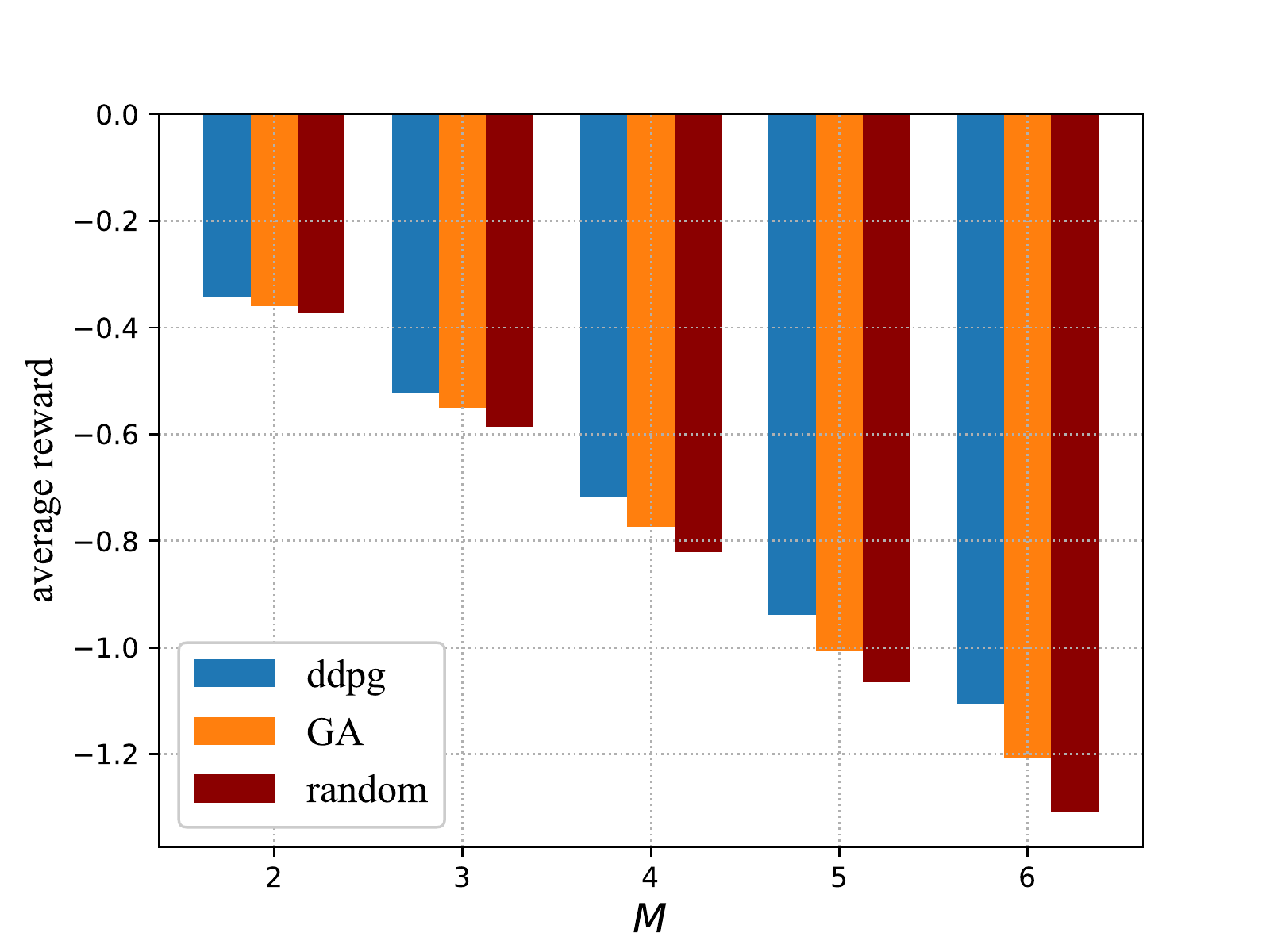}
\caption{Average reward vs. number of devices.}
\label{fig12}
\end{figure}

Fig. \ref{fig11} compares the energy consumption of {three policies}. It can be seen that the energy consumption increases as the number of devices increases. It is because that the interferences will increase the AoI of the system, which imposes the agent to inform the devices to consume more energy to sample more frequently and transmit faster. Moreover, the increasing number of devices contributes to the increasing energy consumption according to Eq. \eqref{eq13}. Meanwhile, the near optimal policy and GA always outperform the random policy, because DDPG and GA can allocate power adaptively to ensure low energy consumption. Moreover, it also can be seen that the near optimal policy always outperforms than GA, that is due to the fact that GA do not have the local optima avoiding scheme like the exploration noise of DDPG. 

\begin{figure}
\centering
\includegraphics[scale=0.5]{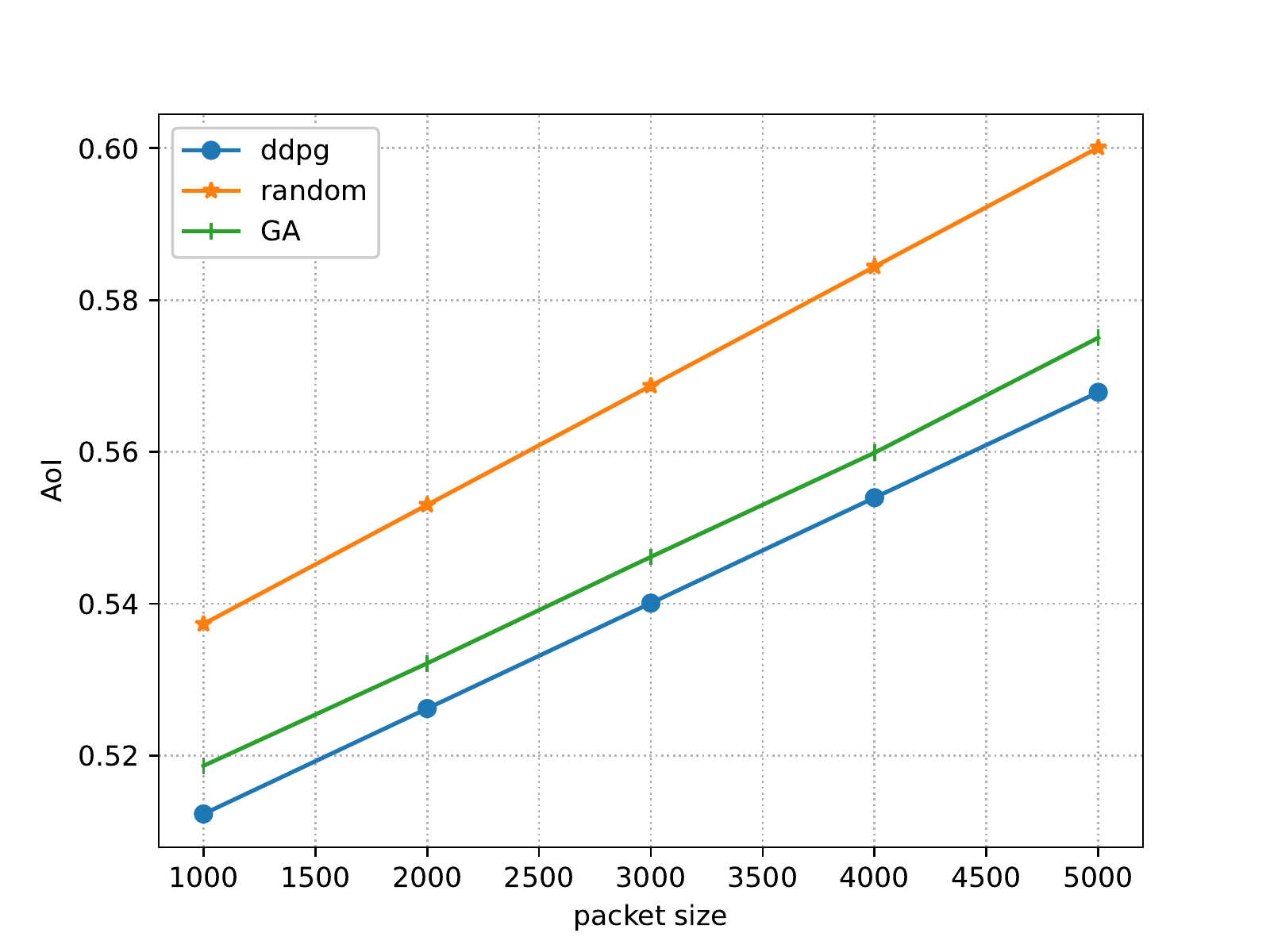}
\caption{AoI of the system vs. packet size.}
\label{fig13}
\end{figure}

\begin{figure}
\centering
\includegraphics[scale=0.5]{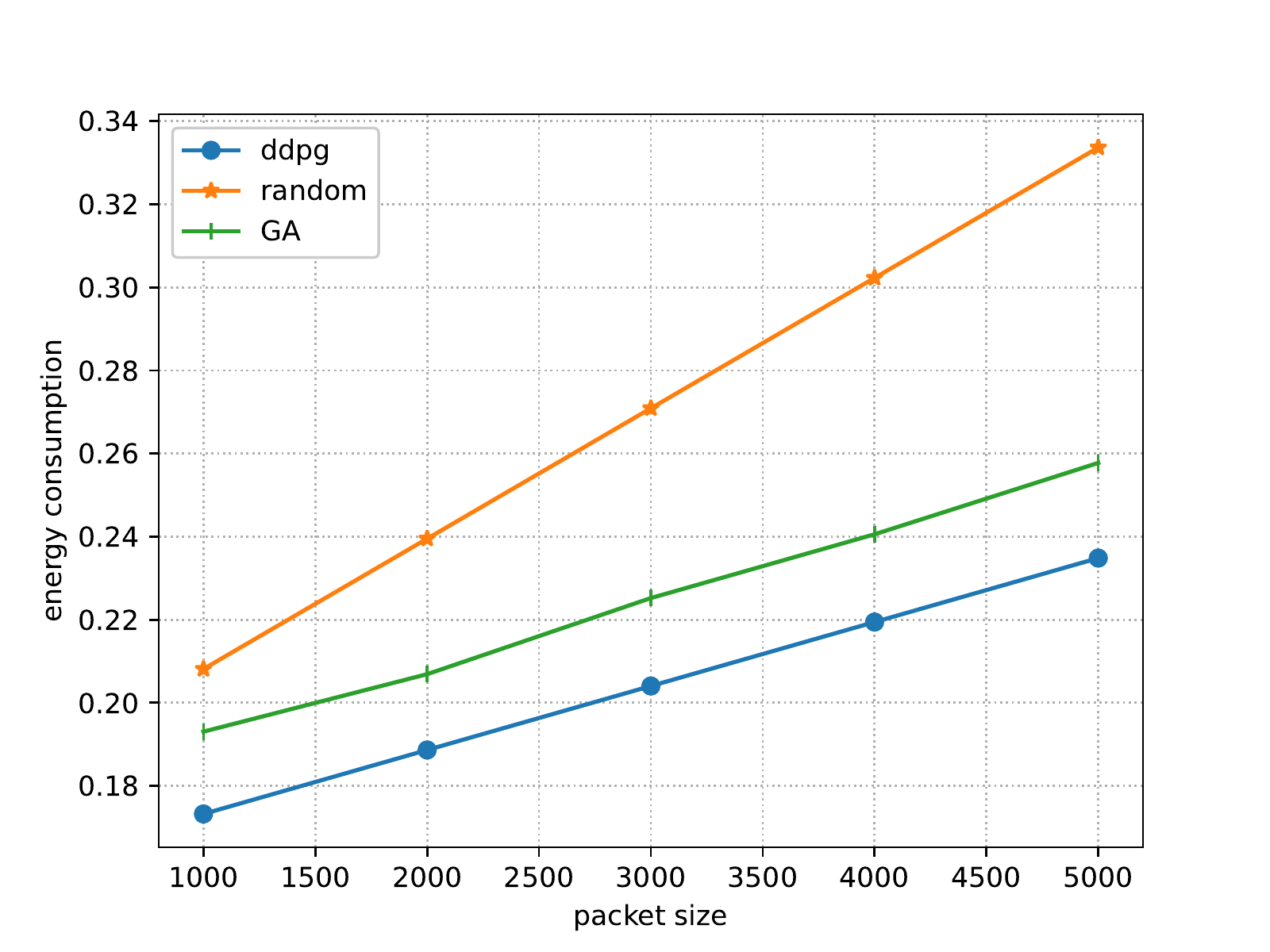}
\caption{Energy consumption vs. packet size.}
\label{fig14}
\end{figure}

Fig. \ref{fig12} compares the average reward under the {three} policies, where the reward is obtained by averaging the test results over all slots. This is shown that the average reward increases as the number of devices increases. It is because that the reward function consists of AoI and energy consumption of the system according to Eq. \eqref{eq19}, and both of them increase as the number of devices increases. Moreover, the average reward under the near optimal policy {and GA} are higher than that of  random policy. It is attributed to the fact that the near optimal policy allocates power according to observed state to maximize the long-term discounted reward{, and the GA obtain the optimal power allocation according to the fitness to maximize the reward. It also can be seen that near optimal policy is always outperforms than GA, it is because that GA aim to find the optimal power allocation based on fitness, i.e., reward in each slot while lose the sight of long-term reward maximization.}

Fig. \ref{fig13} shows the relationship between the AoI of the system and packet size, i.e., $Q$, under {three policies}. It can be seen that the AoI increases as the packet size increases under the three policies. It is because that according to Eq. \eqref{eq10}, the transmission delay is long when the packet size is large, which incurs a large AoI of the system. In addition, we can see that the AoI of the system under the near optimal policy and GA are lower than the AoI under the random policy. It is because the near optimal policy can adjust the power allocation based on the observed state, {and GA obtain optimal power allocation according to fitness, } which can significantly reduce the AoI of the system. {The gap between near optimal power allocation and GA is caused by the local optima of GA.}

Fig. \ref{fig14} shows the relationship between the energy consumption of the system and packet size under {three policies}. It can be seen that the energy consumption increases for the {three} polices when the packet size increases. As shown in Fig. \ref{fig13}, the transmission delay is long when the packet size is large, thus incurring the increase of energy consumption of the system. We can see also that the energy consumption under the near optimal policy {and GA are} lower than that of  random policy. It is because that the near optimal policy can adaptively allocate power {and GA can obtain optimal power allocation according to fitness} to ensure a lower energy consumption. {However, GA may converge into local optima, thus the energy consumption of GA is lower than the near optimal policy obtained by DDPG.}

\section{Conclusions}
\label{sec7}
In this paper, we formulated a problem to minimize the AoI and energy consumption of the MIMO-NOMA IoT system. To solve it, we simplified the formulated problem and proposed the power allocation scheme based on DDPG to maximize the long-term discounted reward. Extensive simulations have demonstrated the superiority of the near optimal policy as compared with the baseline policy.
According to the theoretical analysis and simulation results, we have obtained the following conclusions:

\begin{itemize}
\item  A large number of devices would cause a high AoI of the system. To ensure a lower AoI, the agent will inform the devices to consume more energy to sample more frequently and transmit faster, thus incurring the increase of energy consumption.

\item A large packet size will lead to long transmission delay, which will further cause high AoI of the system. In order to reduce the AoI of the system, the agent would inform the devices to consume more energy to sample more frequently and transmit faster.

\item The near optimal policy trained by DDPG outperforms the baseline policy under different number of users and packet sizes, which has a good capability to suit the system dynamic variation.
\end{itemize}

\bibliographystyle{IEEEtran}
\bibliography{IEEEabrv, mybibfile}


 





\end{document}